\documentclass[epj]{webofc}
\usepackage[utf8]{inputenc}
\usepackage[varg]{txfonts}   
\usepackage{booktabs}
\usepackage{xcolor}
\RequirePackage[colorinlistoftodos,prependcaption,textsize=tiny]{todonotes} 
\definecolor{darkred}{rgb}{0.4,0.0,0.0}
\definecolor{darkgreen}{rgb}{0.0,0.4,0.0}
\definecolor{darkblue}{rgb}{0.0,0.0,0.4}
\usepackage[bookmarks,linktocpage,colorlinks,
   linkcolor = darkred,
   urlcolor  = darkblue,
   citecolor = darkgreen]{hyperref}
\usepackage{comment}
\usepackage{subfigure}
\usepackage{microtype} 
\usepackage{nicefrac} 
\usepackage{amsmath}
\usepackage{epstopdf}
\wocname{EPJ Web of Conferences}
\woctitle{Lattice2017}
\usepackage{array}
\usepackage{tikz}
\usepackage{pgfplots}
\pgfplotsset{compat=newest}

\usetikzlibrary{arrows.meta}
\usetikzlibrary{math}
\usetikzlibrary{calc}

\usepgfplotslibrary{groupplots} 
\usetikzlibrary{pgfplots.groupplots}

\begin{document}
%
\selectlanguage{english}
\title{%
Extracting the Single-Particle Gap in Carbon Nanotubes with \\
Lattice Quantum Monte Carlo
}
\author{%
\firstname{Evan}        \lastname{Berkowitz}    \inst{1}    \and
\firstname{Christopher} \lastname{K\"orber}     \inst{1}    \and
\firstname{Stefan}      \lastname{Krieg}        \inst{2}    \and
\firstname{Peter}       \lastname{Labus}        \inst{3}    \and
\firstname{Timo A.}        \lastname{L\"ahde}      \inst{1}    \and \\
\firstname{Thomas}      \lastname{Luu}          \inst{1}    \fnsep\thanks{Speaker, \email{t.luu@fz-juelich.de}.
Corresponding slides available at \url{https://makondo.ugr.es/event/0/session/101/contribution/319}}
}
\institute{%
Institut f\"ur Kernphysik and Institute for Advanced Simulation, Forschungszentrum J\"ulich, 52425 J\"ulich Germany
\and
J\"ulich Supercomputing Center, Forschungszentrum J\"ulich, 52425 J\"ulich Germany
\and
International School for Advanced Studies, via Bonomea 265, Trieste, Italy
}
\abstract{%
We show how lattice Quantum Monte Carlo simulations can be used to calculate electronic properties of carbon nanotubes in the presence of strong electron-electron correlations.
We employ the path integral formalism and use methods developed within the lattice QCD community for our numerical work and compare our results to empirical data of the Anti-Ferromagnetic Mott Insulating gap in large diameter tubes.
}
\maketitle
\section{Introduction}\label{intro}

Lattice Quantum Monte Carlo methods form an ideal approach for studying non-perturbative phenomena of strongly correlated electrons in low-dimensional crystal systems.
Here the ions of the lattice define the physical, spatial lattice, with the lattice spacing $a$ given by the distance between ions.
As the lattice spacing is typically in the \aa ngstr\"om range ($a=1.42$ \AA\ for graphene), the scale is set and no \emph{spatial} continuum limit is required.
From na\"ive dimensional arguments alone, the low-dimensionality of these systems enhances non-perturbative effects.

In these proceedings we report on progress made in investigating the effects of strong electron correlations in a (quasi) 1-D lattice:  carbon nanotubes.
In particular, we employ standard lattice QCD methods to present calculations of the Anti-Ferromagnetic Mott Insulating (AFMI) gap for tubes of various diameters.
As these tubes have effectively reduced dimensionality (they can be viewed as rolled-up sheets of graphene), the effects of strong correlations are pronounced and both theoretical findings \cite{Balents1997} and empirical data confirm this finding \cite{Deshpande02012009,Bockrath2012,NanotubeCorr}.
We compare our results to empirical data.

\section{Nanotube system}\label{sect:tube description}
\begin{figure}
\center
\begin{tikzpicture} [
  scale=1.00, 
  ]

  \tikzset{
    arrow/.style={
      very thick,
      -{Stealth[length=4mm, width=2mm]},
    }
  }

  \definecolor{CB10orange}{RGB}{255,128,14}
  \definecolor{CB10blue}{RGB}{0,107,164}

  \tikzmath
  {
    \side = .85; 
    \dotradius = 0.10; 
    \imax = 3; 
    \jmax = 4; 
    \startTX = 0; \startTY = 0;
    \endTX = 0; \endTY = 3;
    \endChX = 3; \endChY = 0;
    \endRecX = 3; \endRecY = 3;
    \posAX = 2; \posAY = 2;
    \endAoneX = 2; \endAoneY = 2;
    \endAtwoX = 2; \endAtwoY = 1;
  }

  \node[inner sep=0pt] (startT) at
  ( \side * 3 * \startTX + \side * 3 * cos{60}, \side * 2 * sin{60} * \startTY + \side * sin{60} ) {};

  \node[inner sep=0pt] (endT) at
  ( \side * 3 * \endTX + \side * 3 * cos{60}, \side * 2 * sin{60} * \endTY + \side * sin{60} ) {};

  \node[inner sep=0pt] (endCh) at
  ( \side * 3 * \endChX + \side * 3 * cos{60}, \side * 2 * sin{60} * \endChY + \side * sin{60} ) {};

  \node[inner sep=0pt] (endRec) at
  ( \side * 3 * \endRecX + \side * 3 * cos{60}, \side * 2 * sin{60} * \endRecY + \side * sin{60} ) {};

  \node[inner sep=0pt] (startA) at ( \side * 3 * \posAX, \side * 2 * sin{60} * \posAY ) {};
  \node[inner sep=0pt] (endAone) at
  ( \side * 3 * \endAoneX + \side * 3 * cos{60}, \side * 2 * sin{60} * \endAoneY + \side * sin{60} ) {};
  \node[inner sep=0pt] (endAtwo) at
  ( \side * 3 * \endAtwoX + \side * 3 * cos{60}, \side * 2 * sin{60} * \endAtwoY + \side * sin{60} ) {};

  \clip
  (0.5*\side, 0.5*\side) rectangle (11.0*\side, 6.5*\side);

  \foreach \i in {0,...,\imax}
  {
    \foreach \j in {0,...,\jmax}
    {
      \node[inner sep=0pt]
      (centerA) at ( \side * 3 * \i , \side * 2 * sin{60} * \j ) {};
      \node[inner sep=0pt]
      (centerB) at ( \side * 3 * \i + \side * 3 * cos{60}, \side * 2 * sin{60} * \j + \side * sin{60} ) {};

      \foreach \a in {0,120,-120}
      {
        \draw[dashed] (centerA) -- +(\a:\side);
        \draw[dashed] (centerB) -- +(\a:\side);
      }

      \fill[CB10blue] (centerA) circle (\dotradius);
      \fill[CB10blue] (centerB) circle (\dotradius);
    }
  }

  \foreach \i in {0,...,\imax}
  {
    \foreach \j in {0,...,\jmax}
    {
      \node[inner sep=0pt]
      (centerA) at ( \side * 3 * \i , \side * 2 * sin{60} * \j ) {};
      \node[inner sep=0pt]
      (centerB) at ( \side * 3 * \i + \side * 3 * cos{60}, \side * 2 * sin{60} * \j + \side * sin{60} ) {};

      \foreach \a in {0,120,-120}
      {
        \fill[CB10orange] (centerA) ++(\a:\side) circle (\dotradius);
      }
    }
  }


  \draw[arrow] (startA) -- (endAone) node[pos=0.25, above, inner sep=2mm]
  {$\hspace{1pt} \vec a_1$};
  \draw[arrow] (startA) -- (endAtwo) node[pos=0.25, below, inner sep=2mm]
  {$\hspace{1pt} \vec a_2$};

  \draw[arrow] (startT) -- (endT) node[pos=0.88, right]
  {$3 \hspace{1pt} \vec T$};
  \draw[arrow] (startT) -- (endCh) node[pos=0.91, above]
  {$\vec C_h$};

  \draw[very thick, dashed] (endT) -- (endRec) -- (endCh);

\end{tikzpicture}
\includegraphics[width=.31\columnwidth]{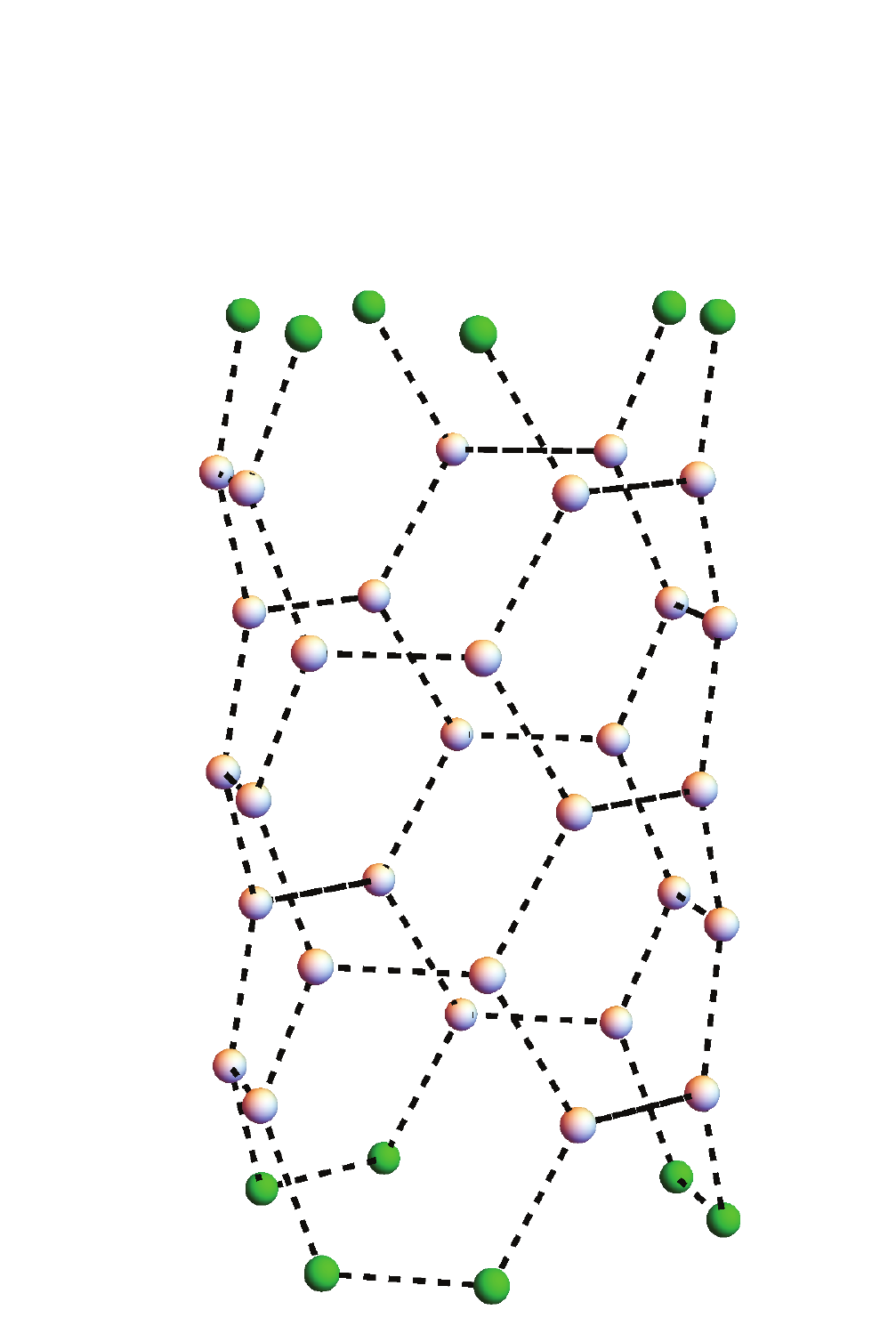}
\caption{
(Left) The basis vectors $\vec{a}_1$ and $\vec{a}_2$ and an example $\vec T$ and $\vec {C}_h$ vectors used to cut the hexagonal sheet for the (3,3) geometry.
(Right) The corresponding tube after subsequent rolling of sheet, and periodic boundary conditions are applied at the green lattice sites.
\label{fig:geometry}
}
\end{figure}
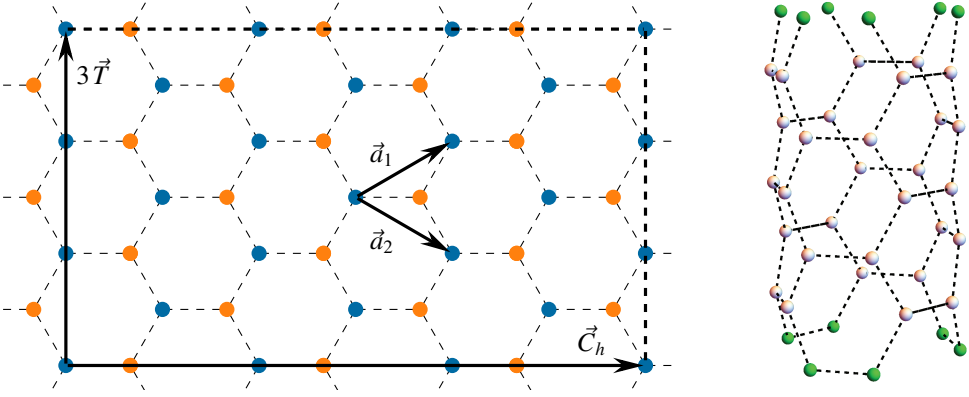
The honeycomb (hexagonal) lattice, \emph{i.e.} graphene, serves as the fundamental structure from which one can construct nanotubes of different types and shapes.
This lattice can be viewed as consisting of two underlying triangular lattices, as shown in the left panel of \autoref{fig:geometry}.
Depending on the geometry of the tube, various properties intrinsic to graphene are inherited by the tube.

\subsection{Geometry}\label{sect:geometry}
Single-wall nanotubes can be viewed as rolled-up sheets of a 2-D honeycomb lattice of carbon ions.
Each tube is characterized by two integers $(n,m)$ (``chirality''), which define the vector $\vec{C}_h=n \vec{a}_1+m \vec{a}_2$ that wraps around the circumference of the tube, and a perpendicular vector $\vec{T}$ whose length gives the minimal tube unit along its axis \cite{Saito1998}.
The left panel of \autoref{fig:geometry} shows these vectors for the 2-D lattice and the right panel the corresponding tube for (3,3) chirality.
For all calculations presented here, we work with $n$=$m$ (``armchair'') geometries and values $n \geq 10$ to minimize systematic errors due to the curvature of the tube.
\subsection{Hamiltonian}\label{sect:hamiltonian}
Because of the localized nature of the $p_z$-orbital electron at each ion, the dynamics of electrons on the lattice is well approximated by nearest neighbor hopping (i.e.
tight-binding), and the effects of electron self-interactions are captured via an onsite interaction, providing the basis for the Hubbard model,
\begin{equation}
\label{eqn:H2}
H=-\kappa\sum_{\langle x,y \rangle}\left(a^\dag_{x}a_{y}^{} + b^\dag_{x}b_{y}^{}\right)
+\frac{U}{2}\sum_{x}\, q_x q_x\ ,
\end{equation}
where
\begin{equation}\label{eqn:charge operator}
q_i=a^\dag_{i}a_{i}^{}-b^\dag_{i}b_{i}
\end{equation}
is the charge operator at ion site $i$, $\kappa$ is the hopping parameter, $U/\kappa$ is the Hubbard ratio, $a^\dag_x$ ($a_x$) is a spin-$\uparrow$ electron creation (annihilation) operator at location $x$, $b^\dag_x$ ($b_x$) is a spin-$\downarrow$ hole creation (annihilation) operator at location $x$ and $\langle x,y\rangle$ indicates sums over all nearest neighbors.
That electrons repel one another provides a basis for the assumption $U\ge0$.
This Hamiltonian represents the electrically neutral, half-filling system, where the average number of electrons per ion site is one.

\subsection{Non-interacting dispersion}\label{sect:non interacting}
\begin{figure}
 \begin{center}
 \includegraphics[width=.7\columnwidth,angle=0]{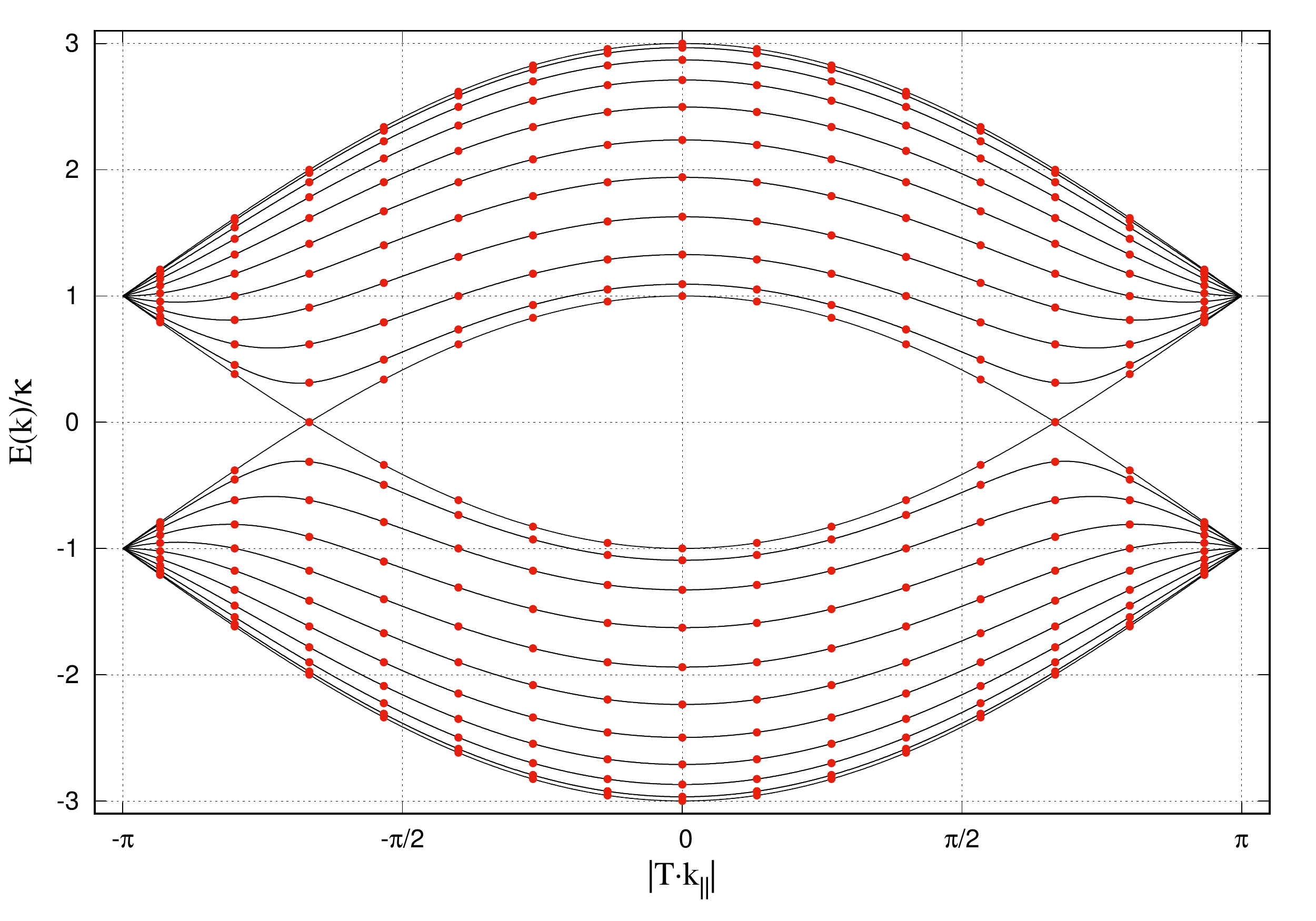}
 \end{center}
\caption{The non-interacting dispersion for a $(10,10)$ armchair tube.
The black lines correspond to a tube of infinite length, while red dots are for a tube of finite length  $L=15|\vec T|$.
Positive energies correspond to electrons, while negative energies correspond to holes.
Note the two Dirac points where the dispersion for electrons and holes coincide.
\label{fig:ni dispersion}}
\end{figure}

In the non-interacting case of $U=0$, the dispersion relation for a single particle can be determined analytically \cite{Saito1998,Charlier:2007zz}.
Because of the finite length around the circumference of the tube, the dispersion exhibits discrete bands.
If the tube is assumed to be infinitely long, then the bands themselves are continuous lines.
In practical calculations, the tube is of finite length with periodic boundary conditions applied at its ends (see right panel of \autoref{fig:geometry}).
The dispersion relation in this case will consist of discrete points along these bands.
\autoref{fig:ni dispersion} shows a sample dispersion for the $(10,10)$ tube.

All armchair configurations $(n,n)$ exhibit two isolated points within the first Brillioun zone, called \emph{Dirac points},  where the dispersion for filled (holes) and unfilled (electrons) bands intersect at $E=0$\footnote{Tubes of finite length $L$ must have $L/|\vec{T}|$ as a multiple of 3 to have access to the Dirac points.}, indicating \emph{metallic} behavior.
Around these Dirac points the dispersion is linear and therefore relativistic, though the characteristic speed is not that of light.
In the presence of electron interactions (i.e.
$U\ne 0$) a gapped anti-ferromagnetic Mott \emph{insulating} (AFMI) state may be formed, separating the bands and removing the Dirac points.
The Dirac points occur at non-trivial momenta $|\vec T\cdot \vec k|=\left(\pm\frac{2\pi}{3},\frac{2\pi}{\sqrt{3}}\right)$ and appropriate momentum projection is required to access information there.

\section{Monte Carlo formulation}\label{sect:MC}
The treatment of Eq.~\eqref{eqn:H2} using the path integral formalism with inverse temperature $\beta$ has been presented elsewhere \cite{Brower:2012zd,Buividovich:2012nx,Smith:2014tha,Luu:2015gpl}.
Here we give a cursory description, concentrating on novel aspects relevant to our work.

\subsection{Setting up the Path Integral}\label{sect:PI}
We consider the expression $e^{-\beta H}$ split into $N_t$ ``time slices'' according to
\begin{equation}
\label{eqn:factor}
e^{-\beta H} \equiv e^{-\delta H}e^{-\delta H}\cdot\cdot\cdot e^{-\delta H}\quad,\quad\delta\equiv\beta/N_t\ .
\end{equation}
We insert a complete set of fermionic states $\psi$ (for electrons) and $\eta$ (for holes) between each exponential to obtain the partition function
\begin{multline}
\label{eqn:Z function}
Z=\text{Tr}\left[e^{-\beta H}\right] = \\
\int \prod_{t=0}^{N_t-1}\left\{\left[\prod_\alpha d\psi^*_{\alpha,t} d\psi_{\alpha,t} d\eta^*_{\alpha,t} d\eta_{\alpha,t}\right]
e^{-\sum_\alpha(\psi^*_{\alpha,t+1}\psi_{\alpha,t+1}^{}+\eta^*_{\alpha,t+1}\eta_{\alpha,t+1}^{})}\langle \psi_{t+1},\eta_{t+1}|e^{-\delta H}|\psi_t,\eta_t\rangle\right\},
\end{multline}
with the condition that $\psi_{N_t}=-\psi_0$ and $\eta_{N_t}=-\eta_0$ (\emph{i.e.}
anti-periodic temporal boundary conditions).
Within each matrix element we then introduce an auxiliary field via the Hubbard-Stratonovich transformation, giving
\begin{multline}
\label{eqn:HS}
\langle \psi_{t+1},\eta_{t+1}|e^{-\delta H}|\psi_t,\eta_t\rangle=
\langle \psi_{t+1},\eta_{t+1}|e^{\delta \kappa\sum_{\langle x,y \rangle}\left(a^\dag_{x}a_{y}+b^\dag_{x}b_{y}\right)-\frac{\delta U}{2}\sum_{x}q_xq_x}|\psi_t,\eta_t\rangle\\
\propto\int \prod_x d\tilde{\phi}_x \langle \psi_{t+1},\eta_{t+1}|e^{\tilde{\kappa}\sum_{\langle x,y \rangle}\left(a^\dag_{x}a_{y}+b^\dag_{x}b_{y}\right)-\frac{1}{2\tilde{U}}
\sum_{x}\tilde{\phi}_x^2 +\sum_x i\tilde{\phi}_x q_x}|\psi_t,\eta_t\rangle,
\end{multline}
where we have introduced the dimensionless variables
\begin{displaymath}
\tilde{\kappa} \equiv \delta \kappa,\quad
\tilde{U} \equiv \delta U,\quad
\tilde{\phi} \equiv \delta \phi\ .
\end{displaymath}
Once standard rules for evaluating matrix elements of normal-ordered operators with fermionic fields are implemented, the integration over fermionic degrees of freedom can be formally done to give
\begin{equation}
\label{eqn:Z function 3}
Z=\int \mathcal{D}\tilde{\phi}\det\left[M(\tilde{\phi})M^\dag(\tilde{\phi})\right]\exp\left\{-\frac{1}{2\tilde{U}}\sum_{x,t=0}^{N_t-1}\tilde{\phi}_{x,t}^2 \right\}
\equiv\int \mathcal{D}\tilde{\phi}\ P\left[\tilde{\phi}\right]\ ,
\end{equation}
where $\mathcal{D}\tilde{\phi}$ is a shorthand notation for $\prod_{x,t=0}^{N_t-1} d\tilde{\phi}_{x,t}$.
The functional fermion matrix $M$ is
\begin{equation}\label{eqn:M}
M(x,t;y,t';\tilde{\phi}) \equiv \delta_{x,y}\left(\delta_{t,t'}-e^{i\tilde{\phi}_{x,t'}}\delta_{t-1,t'}\right)-\tilde{\kappa}\,\delta_{\langle x,y\rangle}\delta_{t-1,t'},
\end{equation}
where $\delta_{\langle x,y\rangle}$ is unity if $x$ and $y$ are nearest-neighbor sites, and zero otherwise.
Equation~\eqref{eqn:M} shows a discretization where the forward differencing in time was done at all lattice sites.
Backward differencing would be equally valid.
In our calculations, we employ both forward and backward differencing on the different sublattices \cite{Luu:2015gpl}.

The probability measure  $P\left[\tilde{\phi}\right]$ is positive definite, therefore standard Monte Carlo techniques can be employed to generate an ensemble $\Phi$ of the auxiliary field $\tilde{\phi}$ with probability distribution $P\left[\tilde{\phi}\right]$.
For this work we utilize pseudo-fermion fields to exponentiate the determinant and employ the hybrid Monte Carlo algorithm \cite{Duane:1987de} to generate such distributions \cite{Smith:2014tha,Luu:2015gpl}.
We generate ensembles for different temporal discretizations and tube characteristics.

\subsection{Momentum wall sources}\label{sect:wall sources}
To extract the quasi-particle dispersion, we analyze the long-time behavior of the single-particle correlator,
\begin{equation}\label{eqn:correlator definition}
\langle a_{x'}(\tau')a_x^\dag(\tau)\rangle=\frac{1}{Z}\int \mathcal{D}\tilde{\phi}\ P\left[\tilde{\phi}\right]\ M^{-1}\left[x',\tau';x,\tau;\tilde{\phi}\right]\approx
\frac{1}{N_\text{cfg}}\sum_{\tilde\phi\in \Phi}M^{-1}\left[x',\tau';x,\tau;\tilde{\phi}\right]
\end{equation}
where $N_\text{cfg}$ is the number of configurations in $\Phi$.
The correlator above contains information of the spectrum at all allowed momentum points.
To extract the spectrum at a particular momentum $ \vec{k}\equiv(k_x,k_y)$, we project onto specific momenta at both the source and sink.
We employ momentum wall sources (for each time slice $\tau$) to perform the momentum projection at the source by solving
\begin{equation}\label{eqn: wall source}
M\left[x',\tau';x,\tau;\tilde{\phi}\right]\ \mathcal{M}\left[x',\tau';\vec{k},\tau;\tilde{\phi}\right]=
\sqrt{\frac{2}{N_i}}
\begin{pmatrix}
e^{i\vec{k}\cdot\vec{x}_1}\\
e^{i\vec{k}\cdot\vec{x}_2}\\
\vdots\\
e^{i\vec{k}\cdot\vec{x}_N}
\end{pmatrix}\ ,
\end{equation}
where $N_i$ is the total number of lattice sites and $\vec{x}_i$ refer
to sites on
\emph{one of the sublattices only}.
It is easy to see that the solution to $\mathcal{M}$ is
\begin{equation}
\mathcal{M}\left[x',\tau';\vec{k},\tau;\tilde{\phi}\right]=\sqrt{\frac{2}{N_i}}\sum_{i}M^{-1}\left[x',\tau';\vec{x}_i,\tau;\tilde{\phi}\right]e^{i\vec{k}\cdot\vec{x}_i}\equiv M^{-1}\left[x',\tau';\vec{k},\tau;\tilde{\phi}\right]\ .
\end{equation}
Finally, we Fourier transform the remaining coordinates to momentum project at the sink
\begin{equation}
\sqrt{\frac{2}{N_i}}\sum_i e^{-i\vec{k}\cdot\vec{x}'_i}M^{-1}\left[\vec{x}'_i,\tau';\vec{k},\tau;\tilde{\phi}\right]\equiv M^{-1}\left[\vec{k},\tau'-\tau;\tilde{\phi}\right]\ .
\end{equation}
We note that there are subtle aspects involved in performing the momentum
projection because of the two underlying sublattices.
These aspects
are discussed in detail in Ref.~\cite{Luu:2015gpl}.
It remains,
however, that for $\tau'-\tau\equiv t\gg0$ (but $t\ll\beta$), the correlator
$M^{-1}\left[k,t;\tilde{\phi}\right]\propto e^{-E\left(\vec{k}\right)t}$, where
$E\left(\vec{k}\right)$ is the desired quasi-particle energy at momentum
$\vec{k}$.
To obtain the value of the gap, we project the correlator at one of
the Dirac momenta $|\vec T\cdot \vec
k_D|=\left(\pm\frac{2\pi}{3},\frac{2\pi}{\sqrt{3}}\right)$, and analyze the
large time behavior to extract $E\left(\vec k_D\right)$.
The gap is given by
$\Delta = 2 E\left(\vec k_D\right)$.

Because of the low dimensionality of our problems, we calculate wall sources for all possible time slices at the source.
This provides us with the means to equivalently calculate the momentum projection of \emph{all-to-all} correlators.

\begin{table}
 \caption{The various geometries and their parameters investigated in this proceeding.  
Here $N_t$ refers to the number of timesteps, $N_i$ the number of carbon atoms, and $N_u$ the number of unit lengths of the tube. Note that the dimension of each system is given by $N_i\times N_t$.
\label{tab:run parameters}}
 \begin{center}
   \begin{tabular}{c c c c c c c}
     \toprule
     chirality & radius (nm) & $\beta$ (eV$^{-1}$) & $U/\kappa$ & $N_u$ & $N_i$ & $N_t$  \\
     \midrule
      (10,10) & 0.68 &$\left\{\begin{matrix} 4\\ 8\end{matrix}\right.$ & $\left\{\begin{matrix}2.0\\ 2.5\\ 3.0 \end{matrix}\right.$ & $\left\{\begin{matrix}15\\ 18\\ 21\end{matrix}\right.$ &$\left\{\begin{matrix} 600\\720\\840\end{matrix}\right.$ & $\left\{\begin{matrix}64\\ 128\\ 196\end{matrix}\right.$ \\[5pt]
     (15,15) & 1.02 & $\left\{\begin{matrix} 4\\ 8\end{matrix}\right.$ & $\left\{\begin{matrix}1.0\\ 1.5\\ 2.0\\ 2.5\\ 3.0\\ 3.5\\4.0\end{matrix}\right.$ & $\left\{\begin{matrix}15\\ 18\\ 21\end{matrix}\right.$ &$\left\{\begin{matrix} 900\\1080\\1260\end{matrix}\right.$ & $\left\{\begin{matrix}64\\ 128\\ 196 \\ 256\end{matrix}\right.$ \\[5pt]
     \bottomrule
   \end{tabular}
 \end{center}
\end{table}

\section{Results}\label{sect:results}
In \autoref{tab:run parameters} we enumerate the types of systems and their parameters that we have investigated.
In principle, our calculations with different $L=N_u |\vec{T}|$ (tube length) allow us to perform an infinite \emph{length} extrapolation, while our multiple $N_t$ runs allow us to perform a continuum limit extrapolation in time.
For weak $U/\kappa\lesssim 1$, we found that our results are nearly constant in $L$, suggesting that $N_u$ is large enough such that our results are already near their infinite volume limits.
For larger couplings we expect the dependence on length to be more pronounced, and we are currently investigating this behavior.  

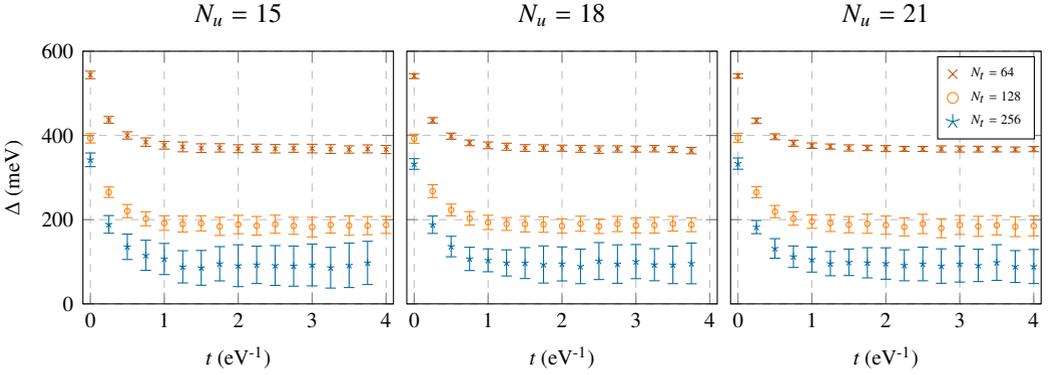
\begin{figure}
 \begin{center}
   \begin{tikzpicture}

  \definecolor{CB10orange}{RGB}{255,128,14}
  \definecolor{CB10blue}{RGB}{0,107,164}
  \definecolor{CB10brown}{RGB}{200,82,0}

  \begin{groupplot}[
    group style={group size=3 by 1, horizontal sep=5pt},
    width=0.40\linewidth,
    xmin=-0.1, xmax=4.1,
    ymin=0, ymax=600,
    grid=major,
    grid style=dashed,
    legend cell align=left,
    legend style={font=\tiny},
    legend image post style={scale=2.0},
    label style={font=\footnotesize},
    tick label style={font=\footnotesize},
    ]

    \nextgroupplot[ 
    xlabel={$t$ (eV\textsuperscript{-1})},
    ylabel={$\Delta$ (meV)},
    title={$N_u=15$},
    ]
    \addplot[
      CB10brown,
      only marks,
      mark=x,
      mark options={mark size=1.2pt},
      error bars/.cd,
      y dir = both,
      y explicit,
      ]
      table [
      x index = {0},
      y index = {1},
      y error index = {2},
      ]
      {./effM.2.dirac_u15_nt64.dat};

    \addplot[
      CB10orange,
      only marks,
      mark=o,
      mark options={mark size=1.0pt},
      error bars/.cd,
      y dir = both,
      y explicit,
      ]
      table [
      x index = {0},
      y index = {1},
      y error index = {2},
      ]
      {./effM.3.dirac_u15_nt128.dat};

      \addplot[
      CB10blue,
      only marks,
      mark=star,
      mark options={mark size=1.4pt},
      error bars/.cd,
      y dir = both,
      y explicit,
      ]
      table [
      x index = {0},
      y index = {1},
      y error index = {2},
      ]
      {./effM.4.dirac_u15_nt256.dat};

      \nextgroupplot[ 
      xlabel={$t$ (eV\textsuperscript{-1})},
      yticklabels={},
      title={$N_u=18$},
      ]

      \addplot[
      CB10brown,
      only marks,
      mark=x,
      mark options={mark size=1.2pt},
      error bars/.cd,
      y dir = both,
      y explicit,
      ]
      table [
      x index = {0},
      y index = {1},
      y error index = {2},
      ]
      {./effM.2.dirac_u18_nt64.dat};

      \addplot[
      CB10orange,
      only marks,
      mark=o,
      mark options={mark size=1.0pt},
      error bars/.cd,
      y dir = both,
      y explicit,
      ]
      table [
      x index = {0},
      y index = {1},
      y error index = {2},
      ]
      {./effM.3.dirac_u18_nt128.dat};

      \addplot[
      CB10blue,
      only marks,
      mark=star,
      mark options={mark size=1.4pt},
      error bars/.cd,
      y dir = both,
      y explicit,
      ]
      table [
      x index = {0},
      y index = {1},
      y error index = {2},
      ]
      {./effM.4.dirac_u18_nt256.dat};

      \nextgroupplot[ 
      xlabel={$t$ (eV\textsuperscript{-1})},
      yticklabels={},
      title={$N_u=21$},
      ]

      \addplot[
      CB10brown,
      only marks,
      mark=x,
      mark options={mark size=1.2pt},
      error bars/.cd,
      y dir = both,
      y explicit,
      ]
      table [
      x index = {0},
      y index = {1},
      y error index = {2},
      ]
      {./effM.2.dirac_u21_nt64.dat};

      \addplot[
      CB10orange,
      only marks,
      mark=o,
      mark options={mark size=1.0pt},
      error bars/.cd,
      y dir = both,
      y explicit,
      ]
      table [
      x index = {0},
      y index = {1},
      y error index = {2},
      ]
      {./effM.3.dirac_u21_nt128.dat};

      \addplot[
      CB10blue,
      only marks,
      mark=star,
      mark options={mark size=1.4pt},
      error bars/.cd,
      y dir = both,
      y explicit,
      ]
      table [
      x index = {0},
      y index = {1},
      y error index = {2},
      ]
      {./effM.4.dirac_u21_nt256.dat};

      \legend{
        \, $N_t = 64$,
        \, $N_t = 128$,
        \, $N_t = 256$,
      }

    \end{groupplot}
  \end{tikzpicture}
 \end{center}
 \vspace*{-5mm}
 \caption{
   Value of the gap determined from the effective mass at the Dirac point with coupling
   $U/\kappa=1$ and $\beta=8$ eV$^{-1}$ for $N_u=15,18,21$ and $N_t=64,128,256$.
   Results are nearly independent in $N_u$ for this (weak) value of the coupling,
   but demonstrate strong dependence on $N_t$.
 }
 \label{fig:eff mass}
\end{figure}

Regardless of coupling, all our results showed strong dependence on $N_t$, and thus a temporal continuum extrapolation was essential.
\autoref{fig:eff mass} shows these findings.
\begin{figure}
\center
\includegraphics[width=.5\columnwidth]{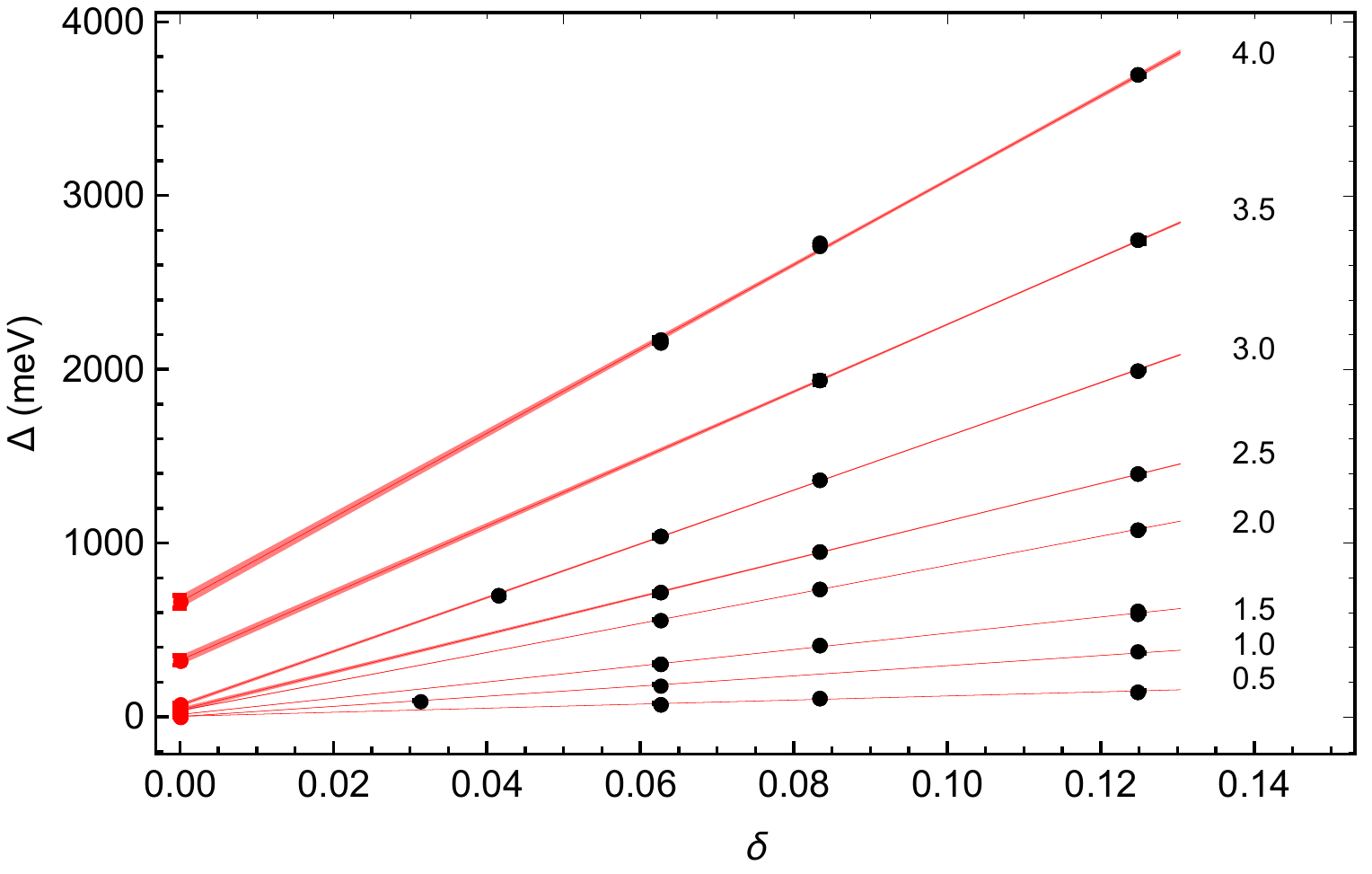}\includegraphics[width=.495\columnwidth]{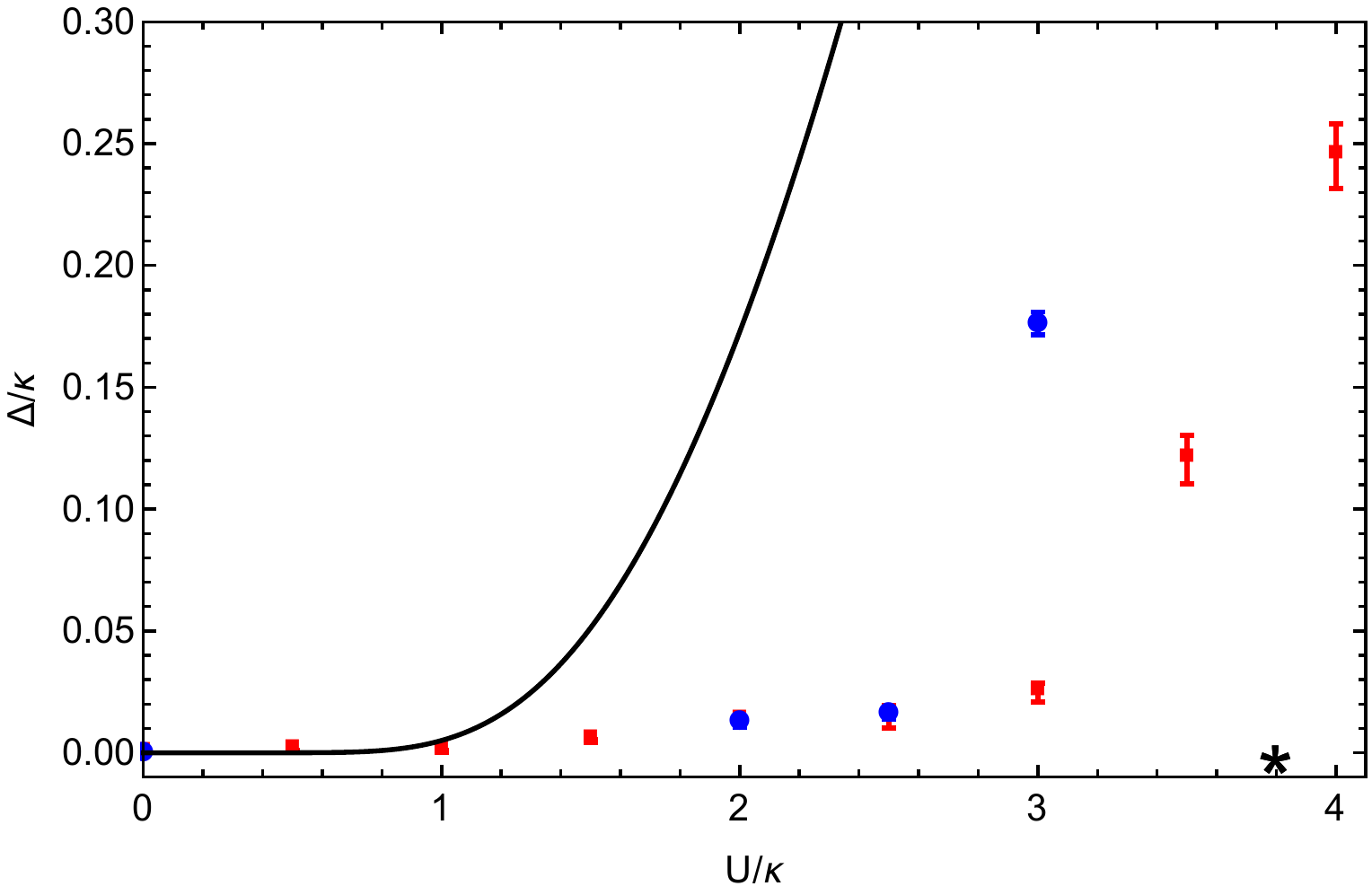}
\caption{(Left) The dependence of the gap $\Delta$ on $\delta=\beta/N_t$ from the (15,15) tube (black points, error bars too small to be seen), and their continuum extrapolation (red points).
The different lines represent calculations with different Hubbard ratios, with the lowest being $U/\kappa=0.5$ and highest $U/\kappa=4$, as labelled in the figure.
(Right) The dependence of the Mott-insulating gap $\Delta$ (in units of $\kappa$) on Hubbard ratio $U/\kappa$ for (15,15) tube (red squares) and (10,10) tube (blue circles).
The solid line shows the limiting 1-D analytical result obtained using the Bethe ansatz \cite{Lieb1968,Lieb2003a}.  Also indicated in the right plot is the location of the critical coupling (black star) for the 2-D hexagonal Hubbard model \cite{Otsuka2016}.  The red points of the two plots coincide with each other.
 \label{fig:continuum limit}}
\end{figure}
\autoref{fig:continuum limit} shows explicit examples of the dependence of the gap $\Delta$ on the timestep $N_t$, and our extrapolated continuum results using the function
\begin{equation}\label{eqn:linear}
\Delta(\delta)=\Delta_0+\alpha\  \delta\ ,
\end{equation}
where $\delta=\beta/N_t$ and $\Delta_0$ is continuum result.
Our
correlation functions are accurate to $\mathcal{O}(\delta^2)$
\cite{negele1988quantum}, and so the calculated effective masses
should scale as $\mathcal{O}(\delta)$, which motivates the use of Eq.~\eqref{eqn:linear}.
We combine our extrapolated points in the right panel of \autoref{fig:continuum limit} and show the dependence of $\Delta/\kappa$ as a function of Hubbard ratio $U/\kappa$ for different tube geometries.
Also shown in the figure for comparison is the analytical result for the 1-D Hubbard model obtained with the Bethe ansatz \cite{Lieb1968,Lieb2003a}.  The 1-D result is the limit in which the tubes have vanishing radii.

\section{Comparison with data and discussions}\label{sect:ta da}
\begin{figure}
  \begin{center}
    \begin{tikzpicture}

  \definecolor{CB10orange}{RGB}{255,128,14}
  \definecolor{CB10blue}{RGB}{0,107,164}
  \definecolor{CB10brown}{RGB}{200,82,0}
  \definecolor{CB10gray}{RGB}{89,89,89}

  \begin{semilogyaxis}[
    width=0.75\linewidth,
    height=0.50\linewidth,
    xlabel={$r$ (nm)},
    ylabel={$\Delta$ (meV)},
    xmin=0.5, xmax=4.1,
    xtick={1,1.5,...,4.0},
    ymin=10, ymax=1100,
    ytick={10,100,1000},
    yticklabels={10,100,1000},
    major tick length=3.5pt,
    legend cell align=left,
    legend image post style={scale=1.5},
    ]

    \addplot[
      CB10brown,
      only marks,
      mark=square*,
      mark options={mark size=1.2pt},
      error bars/.cd,
      y dir = both,
      y explicit,
      ]
      table [
      x index = {0},
      y index = {1},
      ]
      {./exp.dat};

    \addplot[
      CB10blue,
      only marks,
      mark=o,
      mark options={mark size=1.2pt},
      error bars/.cd,
      y dir = both,
      y explicit,
      ] table [ x index = {0}, y index = {1}, y error index = {2}, ] {./gapU2.dat};

    \addplot[
      CB10orange,
      only marks,
      mark=*,
      mark options={mark size=1.2pt},
      error bars/.cd,
      y dir = both,
      y explicit,
      ] table [ x index = {0}, y index = {1}, y error index = {2}, ] {./gapU3.dat};

    \addplot[
      CB10gray,
      only marks,
      mark=triangle,
      mark options={mark size=1.2pt},
      error bars/.cd,
      y dir = both,
      y explicit,
      ] table [ x index = {0}, y index = {1}, y error index = {2}, ] {./gapU35.dat};

      \legend{
        \, experimental,
        \, $\nicefrac{U}{\kappa} = 2.0$,
        \, $\nicefrac{U}{\kappa} = 3.0$,
        \, $\nicefrac{U}{\kappa} = 3.5$,
      }

  \end{semilogyaxis}
\end{tikzpicture}
  \end{center}
  \vspace*{-5mm}
\caption{The dependence of the gap on tube radius $r$ for different values of the Hubbard ratio, with comparison to experimental data extracted from Ref.~\cite{Deshpande02012009}.
 \label{fig:U dependence}}
\end{figure}
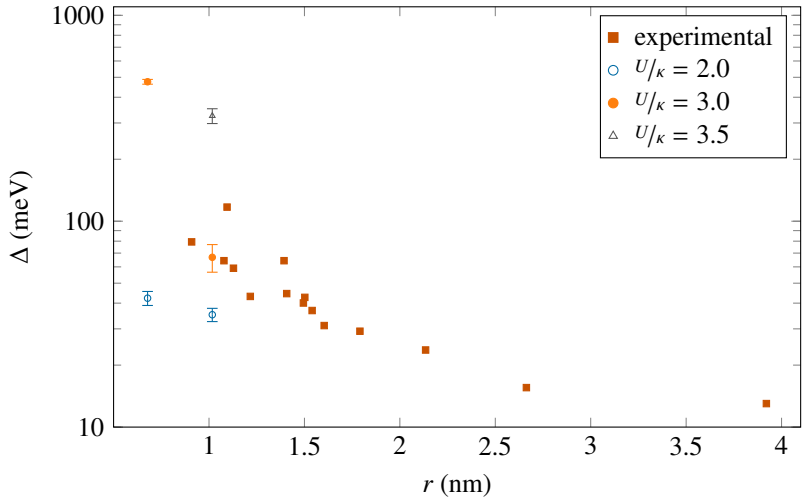
Measurements of ultra-pure carbon nanotubes confirm the presence of a Mott-insulating (AFMI) gap \cite{Deshpande02012009}.
In \autoref{fig:U dependence} we have extracted this data from Ref.~\cite{Deshpande02012009} and plotted it next to our numerical results.
We find that a value of $U/\kappa\approx 3$ is matches the experimental data at the small tube radus of about 1~nm, and it will be interesting to see if this also value reproduces the measurements at larger radii.

In the limit that the chiral index $n\rightarrow\infty$, the radius of our tube goes to infinity and we recover the graphene limit.
Past numerical simulations of graphene suggest that there is a phase transition between semi-metal and Mott insulator that occurs near $U_c/\kappa\approx 3.5$ \cite{Meng2010}. More recent calculations have pushed this value larger, $U_c/\kappa\approx 3.8$ \cite{Otsuka2016}.  Since matching the nanotube data favors a lower $U$, our results suggest that graphene is a semi-metal.

Current work is ongoing to calculate the dispersion of multi-quasiparticle systems, such as the exciton (1-particle/1-hole) and trion (2-particle/1-hole or 2-hole/1-particle) systems.
The latter is quite interesting since experiments show that the system is bound in doped (\emph{not} half-filling) systems \cite{PhysRevLett.106.037404}.
Here the techniques for computing multi-particle interpolating operators in lattice QCD calculations can be directly applied to calculate the spectrum of these systems.
The ability to simulate doped systems is akin to introducing a chemical potential, which will introduce a complex phase in $P\left[\phi\right]$ leading to sign problems.
The tube system offers an ideal setting for developing and investigating novel sampling techniques (see, \emph{e.g.}, Ref.~\cite{bedaque419}) to tackle the sign problem due to its small dimensionality and quick calculational turn-around time.


\end{document}